\documentclass[aps,pra,twocolumn,floatfix]{revtex4-2}
\usepackage{amsmath, amssymb, amsthm}
\usepackage{graphicx}
\usepackage{hyperref}
\usepackage{multirow}
\usepackage{float}
\usepackage{subcaption}
\usepackage{tikz}
\usepackage{pgfplots}
\usepackage{booktabs}
\usepackage{dblfloatfix}
\usepackage{afterpage}
\usepackage{bookmark}
\IfFileExists{soul.sty}{\usepackage{soul}}{}
\usepackage{xcolor}
\usepackage{threeparttable}
\usepackage{booktabs}
\usepackage{ragged2e}
\makeatletter
\long\def\@makecaption#1#2{%
  \vskip\abovecaptionskip
  \sbox\@tempboxa{#1: #2}%
  \ifdim \wd\@tempboxa >\hsize
    {\justifying #1: #2\par}%
  \else
    {\justifying #1: #2}%
  \fi
  \vskip\belowcaptionskip
}
\makeatother
\begin{document} 
\title{Super-Link Fragility in Asymmetric W-Class States under Quantum Noise}

\author{Sougata Bhattacharyya$^{1}$, Fatih Ozaydin$^{2,3,4}$, Sovik Roy$^{5}$\vspace{0.25cm}}

\affiliation{%
$^{1}$Department of Astronomy, Astrophysics and Space Engineering, Indian Institute of Technology, Indore 453552, India\vspace{0.2cm}\\
$^{2}$Department of Data Science and AI, Tokyo International University, 4-42-31 Higashi-Ikebukuro, Toshima-ku, Tokyo 170-0013, Japan\vspace{0.2cm}\\
$^{3}$Institute for International Strategy and Emerging Technologies, Tokyo International University, 4-42-31 Higashi-Ikebukuro, Toshima-ku, Tokyo 170-0013, Japan\vspace{0.2cm}\\
$^{4}$Nanoelectronics Research Center, Kosuyolu Mah., Lambaci Sok., Kosuyolu Sit., No.\ 9E/3 Kadikoy, İstanbul 34718, Türkiye\vspace{0.2cm}\\
$^{5}$Department of Mathematics, Techno Main Salt Lake, Techno India Group, EM 4/1, Sector V, Kolkata 700091, India%
}

\date{\today}

\begin{abstract}
	The asymmetric three-qubit $W$-class state $|\overline{W_3^L}\rangle$ defines an isosceles entanglement-network geometry, (a) two vertex-base ($VB$) links form stronger bipartite connections, (b) while the base-base ($BB$) link is weaker. This suggests that concentrating entanglement into a \textit{super-link} may be advantageous for quantum-network tasks. Here, we show that this intuition is incomplete. We analytically compare the bipartite concurrence dynamics of the symmetric $|W\rangle$ state and the asymmetric $|\overline{W_3^L}\rangle$ state, which differ both in entanglement-network geometry and excitation sector under standard noise models. In the absence of noise, the concurrence hierarchy is $\mathcal{C}_{VB} > \mathcal{C}_W > \mathcal{C}_{BB}$. Under phase damping, this hierarchy is preserved for all noise strengths and no entanglement sudden death occurs. Under amplitude damping, however, the hierarchy is reordered. The symmetric $|W\rangle$ state becomes the most robust, while the base-base concurrence of $|\overline{W_3^L}\rangle$ vanishes at the finite threshold of parameter $\gamma$. We term this reordering as the \textit{Super-Link Fragility Effect}. The same structural asymmetry that produces a stronger vertex-base link also makes it more vulnerable to energy dissipation when coupled with multi-excitation amplitudes. Under depolarization, the asymmetry advantage is erased, with $\mathcal{C}_W$ and $\mathcal{C}_{VB}$ sharing the same sudden-death threshold for some value of the parameter $p$, while $\mathcal{C}_{BB}$ disappears earlier at some other value of the parameter $p$. The generalized amplitude damping channel continuously connects the damping-dominated regime to the pure-excitation limit, where the initial hierarchy is restored. These results show that entanglement robustness in $W$-class resources is controlled not by initial concurrence alone, but by the joint structure of entanglement-network geometry, excitation sector, and noise symmetry.
	%\textbf{Keywords:} Super-Link Fragility, Entanglement-network geometry, $W$ class states.
\end{abstract}

\maketitle

\section{Introduction}
\label{sec:intro}

Quantum entanglement is a foundational resource for quantum information
processing, enabling quantum communication, computation, and sensing
\cite{Horodecki2009,Nielsen2010,Bennett1993,Gisin2007,Ladd2010}.
Among multipartite entangled states, the symmetric three-qubit $|W\rangle$
state occupies a privileged position. It distributes its single
excitation uniformly across all three qubits, conferring remarkable
robustness against particle loss~\cite{Dur2000,Acin2001,Verstraete2002}.
This property distinguishes it sharply from the
Greenberger-Horne-Zeilinger (GHZ) state~\cite{Dur2000} and has made
$|W\rangle$ a preferred building block for distributed quantum networks
\cite{Kimble2008,Agrawal2006,Agrawal2002,Haffner2005}.

However, not all $W$-class states are symmetric.
Asymmetric $W$-class states break permutation symmetry and thereby
exhibit unequal bipartite entanglement links. This asymmetry leads to richer physical behavior and introduces new constraints and requirements for practical applications. For example, unlike conventional symmetric $W$ states, asymmetric $W$ states are required for perfect superdense coding~\cite{li2016generating}. Moreover, even the preparation of symmetric $W$ states remains challenging, motivating the development of fusion-based approaches~\cite{bugu2013enhancing, yesilyurt2013optical, ozaydin2014fusing, zang2015generating}, expansion strategies~\cite{yesilyurt2016deterministic, zang2016deterministic, ozaydin2021deterministic}, quantum eraser schemes~\cite{kim2020efficient}, and preparation methods based on Pauli spin blockade~\cite{bugu2020preparing}. Efficient generation of asymmetric $W$ states is even more demanding, since each inequivalent asymmetry class generally requires additional modifications or entirely new preparation protocols.

The state $|\overline{W_3^L}\rangle$,
introduced by Lohmayer
\textit{et al.}~\cite{Lohmayer2006}, exemplifies this asymmetry. It
lies in the \textit{two-excitation} subspace (two qubits in $|1\rangle$,
one in $|0\rangle$), in contrast to the \textit{single-excitation}
$|W\rangle$ state.
This difference in excitation sector, together with the geometric
asymmetry, is central to the decoherence dynamics studied here and will
be made explicit throughout. This state $|\overline{W_3^L}\rangle$ possesses a specific geometry: its \textit{isosceles entanglement-network geometry} means that qubit $A$ (the vertex) is more strongly entangled with each base qubit ($B$ or $C$) (i.e.\ $AB$ or $AC$) than the base qubits are with each other (i.e.\ $BC$).
%\newpage

To quantify this structure, we work with \textit{bipartite concurrence}
\cite{Wootters1998,Hill1997}.
Tracing out one qubit from the three-qubit state yields a two-qubit
mixed state; its concurrence measures the residual bipartite link.
Throughout this work, we shall follow the following terminologies.
By $\mathcal{C}_W$ we denote the concurrence of any bipartite reduction of the symmetric
$|W\rangle$ state (all three pairs $A, ~B,~ \& ~C$ are equivalent by symmetry);
$\mathcal{C}_{VB}$ denotes the concurrence of the vertex-base i.e. $AB ~ \& ~ AC $ pairs
($\mathcal{C}_{AB}=\mathcal{C}_{AC}$) of the asymmetric state; and
$\mathcal{C}_{BB}$ signifies the concurrence of the base-base i.e. $BC$ pair ($\mathcal{C}_{BC}$) of the asymmetric state.

The asymmetry of the network naturally introduces inequivalent entanglement channels between different
pairs of nodes. In particular, the coupling between the vertex-base nodes differs from that of the symmetric configuration, raising the question of how such structural asymmetry influences both the initial distribution of entanglement and its subsequent evolution under decoherence. In this work, we investigate these effects and identify characteristic signatures associated with the asymmetry topology. It has been observed that the stronger vertex-base link, which we call the \textit{super-link}, has a higher initial concurrence than the symmetric state. A natural question then arises, does this initial advantage persist under realistic noise, or does it become a liability?

The answer turns out to depend critically on the noise mechanism.
Different decoherence channels such as (i) phase damping (pure dephasing),
(ii) amplitude damping (energy dissipation)~\cite{Plenio1998},
(iii) depolarization~\cite{Nielsen2010}, and (iv) generalized amplitude damping
(GADC)~\cite{Srikanth2008,Gao2008}, interact with the state's excitation
structure in fundamentally different ways.

Previous works have studied entanglement dynamics of symmetric $W$
states~\cite{Siomau2010,Espoukeh2015,Ozaydin2014,Carvalho2004} and the phenomenon
of Entanglement Sudden Death (ESD) in two-qubit
systems~\cite{Yu2004,Yu2009,Yu2007,Almeida2007,Huang2007,Weinstein2009,Lopez2008,Zyczkowski2001}, but the
channel-dependent hierarchy of bipartite links within an
\textit{asymmetric} $W$-class state has not been systematically
characterized.

In this work we perform a comprehensive analytical study of the
bipartite concurrence dynamics of both states under all four channels.
Our main result is that the \textit{initial hierarchy
$\mathcal{C}_{VB} > \mathcal{C}_W > \mathcal{C}_{BB}$ is not a reliable predictor of robustness}.
Under phase damping the ordering is preserved and no ESD occurs. Under amplitude damping the hierarchy is reordered  ($\mathcal{C}_W > \mathcal{C}_{VB} > \mathcal{C}_{BB}$),
with $\mathcal{C}_{BB}$ exhibiting ESD at $\gamma=1/2$. We call this channel-dependent reordering -  the \textit{Super-Link Fragility Effect}.
Under depolarization the asymmetry advantage is washed out ($\mathcal{C}_{VB}$
and $\mathcal{C}_W$ share sudden-death threshold at $p=3/8$ and $C_{BB}$ disappears earlier at $p=3/10$). The GADC interpolates continuously between these extremes.

The remainder of the paper is organized as follows. In Sec. \ref{sec:preliminaries}, we introduce the quantum states under consideration, outline the concurrence measure used to quantify pairwise entanglement, and briefly review the decoherence channels studied in this work. Sec. \ref{sec:geometry} develops the entanglement-network representation and examines the geometric features arising from structural asymmetry. The effects of environmental noise on the resulting entanglement structure are analyzed in Sec. \ref{sec:decoherence}. In Sec. \ref{sec:synthesis}, we discuss the implications of the observed dynamical behavior and compare the responses of the different correlation channels. Finally, Sec. \ref{sec:conclusion} summarizes the main findings and presents concluding remarks. Detailed derivations are in the Appendices.

\section{Preliminaries}
\label{sec:preliminaries}

\subsection{States}
\label{subsec:states}

The symmetric three-qubit $|W\rangle$ state is defined as~\cite{Dur2000}
\begin{equation}
|W\rangle = \frac{1}{\sqrt{3}}\bigl(|001\rangle + |010\rangle
+ |100\rangle\bigr),
\end{equation}
which lies in the \textit{single-excitation subspace} (one qubit in
$|1\rangle$, two qubits are in $|0\rangle$ state).

On the other hand,  the asymmetric state $|\overline{W_3^L}\rangle$ is defined
as~\cite{Lohmayer2006}
\begin{equation}
|\overline{W_3^L}\rangle = \frac{1}{2}|110\rangle
+ \frac{1}{2}|101\rangle + \frac{1}{\sqrt{2}}|011\rangle,
\end{equation}
which lies in the \textit{two-excitation subspace}.
This difference in excitation number is \textit{not} merely a geometric
detail, rather it fundamentally affects the response to amplitude damping, as
we show in Section~\ref{subsec:amplitude_damping}.

\subsection{Entanglement architecture: Borromean and Hopf Modes}
\label{subsec:geometry_modes}

To understand the entanglement architecture of these states, we map their
bipartite correlations to conceptual link-based representations
\cite{Kauffman2002,Kauffman1999}.
We emphasize that this mapping is \textit{conceptual} and serves as an
intuitive visualization; the actual dynamics are governed by the
density-matrix evolution.
Three-qubit entanglement can be understood as a contextual hybrid of
two primary modes revealed by projective measurements on a single qubit.

%\begin{itemize}
%    \item 
    \textit{Borromean Mode:} Measuring a qubit in the
    $|0\rangle$ basis and finding the residual bipartite state
    completely separable ($\mathcal{C}_{l_1}=0$, where $\mathcal{C}_{l_1}$ is the $l_1$- norm of coherence) is analogous to severing a
    Borromean ring, which causes the remaining two rings to fall
    apart~\cite{aravind1997,Sugita2007,bhattacharyya2026symmetric,bhattacharyya2026entanglement}.

\begin{figure}[!t]
\centering
\includegraphics[width=0.55\columnwidth]{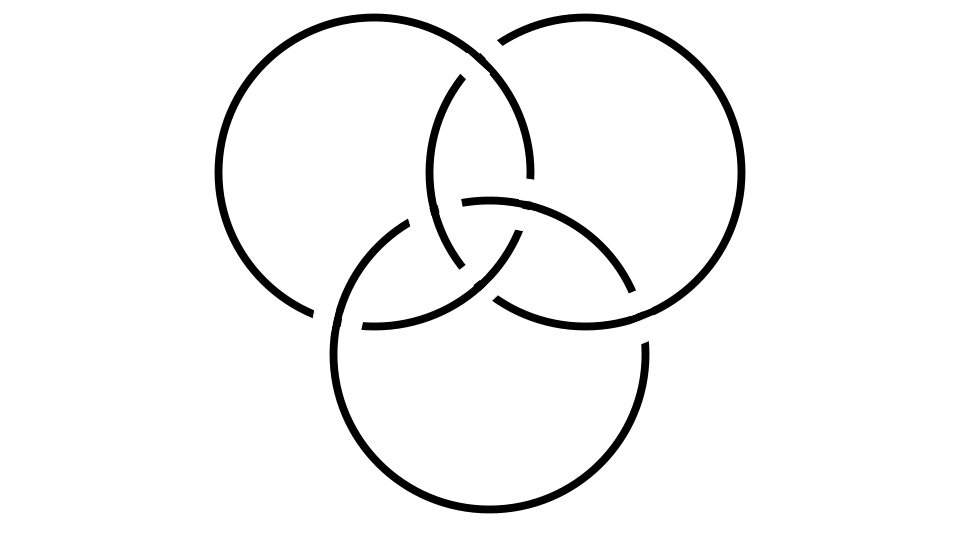}
\caption{\textbf{Borromean Rings.} Three mutually interlocked rings in which no two rings are directly linked as a pair, yet the three together cannot be separated. This topology serves as an analogy for the \textit{Borromean mode} of three-qubit entanglement: when any single qubit is measured and found in $|0\rangle$, the residual two-qubit state of the remaining pair collapses to a fully separable product state with zero coherence ($C_{l_1}=0$), mirroring how severing one Borromean ring causes the other two to fall entirely apart.}
\label{fig:borromean}
\end{figure}

    %\item 
    \textit{Hopf Mode:} Measuring a qubit in the $|1\rangle$
    basis and finding the residual bipartite state entangled
    ($\mathcal{C}_{l_1}>0$) is analogous to a 3-Hopf link, where removing one ring still leaves the other two interlinked \cite{bhattacharyya2026symmetric,bhattacharyya2026entanglement}.

\subsubsection{Coherence and the $|\overline{W_3^L}\rangle$ State}

To quantify residual quantum coherence we use the $l_1$-norm of
coherence~\cite{Baumgratz2014},which is defined as
\begin{equation}
\mathcal{C}_{l_1}(\psi) = \Bigl(\sum_i |c_i|\Bigr)^2 - 1,
\end{equation}
where $c_i$'s are the amplitudes in the computational basis.

The symmetric $|W\rangle$ state has initial coherence
\begin{equation}
\label{coheriniw}
\mathcal{C}_{l_1}^{\text{init}}(|W\rangle)=2.
\end{equation}
Measuring any qubit gives (i) outcome $|0\rangle$ (with probability $\frac{2}{3}$), the residual state is
$\frac{1}{\sqrt{2}}(|01\rangle+|10\rangle)$ and the coherence $\mathcal{C}_{l_1}$ of the residual state is $=1$;
(ii) outcome $|1\rangle$ (with probability $\frac{1}{3}$),  residual state is $|00\rangle$ whose coherence is
$\mathcal{C}_{l_1}=0$.

On the other hand, the asymmetric $|\overline{W_3^L}\rangle$ state has initial coherence
\begin{eqnarray}
\mathcal{C}_{l_1}^{\text{init}}(|\overline{W_3^L}\rangle) &=& \Bigl(1+\tfrac{1}{\sqrt{2}}\Bigr)^2 - 1\nonumber\\
&=& \tfrac{1}{2}+\sqrt{2} \approx 1.914.
\end{eqnarray}
However upon measurement, we get the following.
When the outcome is $|0\rangle$ (Borromean mode): residual state is $|11\rangle$ and coherence is
\begin{eqnarray}
    \label{cohboro}
          \mathcal{C}_{l_1}=0.
\end{eqnarray}

For outcome $|1\rangle$ on vertex $A$ with probability $\frac{1}{2}$ (Hopf mode): residual state is
$\frac{1}{\sqrt{2}}(|10\rangle+|01\rangle)$ and coherence
\begin{eqnarray}
\label{vertexAwbarcoh}
    \mathcal{C}_{l_1}=1.
\end{eqnarray}

For outcome $|1\rangle$ on base $B$ or $C$ with probability $\frac{3}{4}$ (Hopf mode):
residual state is $\frac{1}{\sqrt{3}}|10\rangle+\sqrt{\frac{2}{3}}|01\rangle$ and coherence is given as
\begin{eqnarray}
\label{vertexBwbarcoh}
\mathcal{C}_{l_1} = \Bigl(\frac{1+\sqrt{2}}{\sqrt{3}}\Bigr)^2 - 1
= \frac{2\sqrt{2}}{3} \approx 0.943.
\end{eqnarray}

\begin{figure}[!t]
	\centering
	\includegraphics[width=0.55\columnwidth]{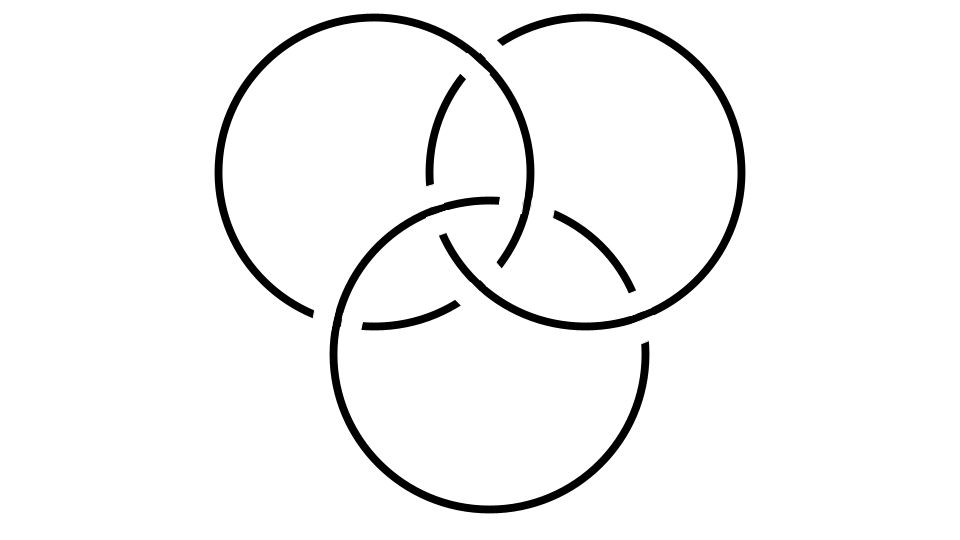}
	\caption{\textbf{3-Hopf Link.} Three rings arranged so that removing any single ring still leaves the remaining two interlinked. This topology serves as an analogy for the \textit{Hopf mode} of three-qubit entanglement: when any single qubit is measured and found in $|1\rangle$, the residual two-qubit state retains non-zero coherence \textcolor{blue}{($\mathcal{C}_{l_1}>0$)} and bipartite entanglement, mirroring the persistent linkage of the remaining two Hopf rings.}
	\label{fig:hopf}
\end{figure}

From the above, we observe that there is a subtle difference between the $|W\rangle$ and $|\overline{W_3^L}\rangle$ states. In case of $|W\rangle$ state,  measuring qubits with respect to vertices $A,~B, ~C$ yield the same outcome whereas for $|\overline{W_3^L}\rangle$ state, the measurements with respect to vertices $A,~B$ and $C$ are showing different results in terms of coherence, (see Eqs.(\ref{cohboro}), (\ref{vertexAwbarcoh}) and (\ref{vertexBwbarcoh}) where two different values $1$ and $0.943$ are observed. This quantitatively confirms the isosceles
nature of the entanglement-network geometry: the vertex qubit leaves
behind a more robustly coherent resource than the base qubits do. In simple words, vertex $A$ is the privileged hub.

\begin{figure}[!ht]
\centering
\includegraphics[width=\columnwidth]{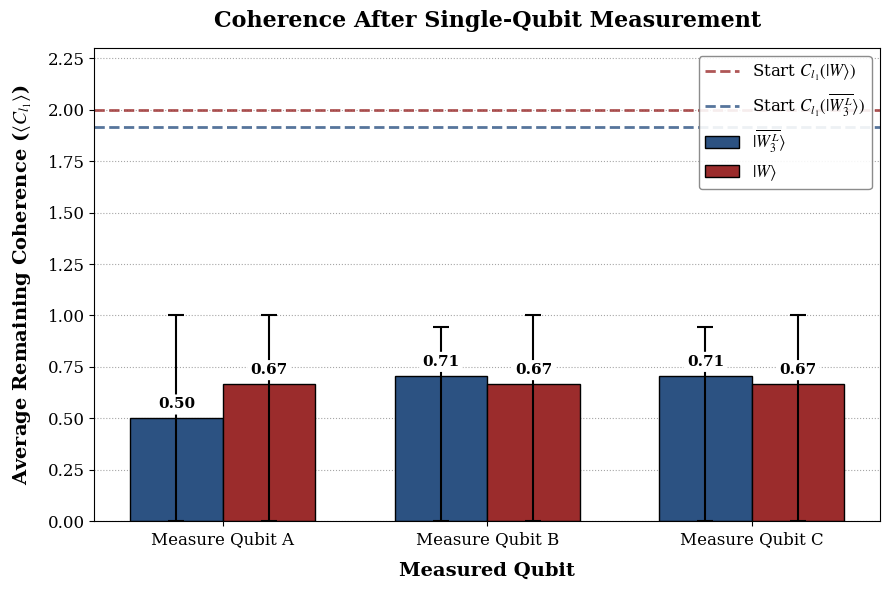}
\caption{\textbf{$l_1$-norm coherence of the two-qubit remainder after
single-qubit measurement.}
For each measured qubit ($A$, $B$, or $C$), the two endpoints of the
error bar mark the two possible coherence values corresponding to the
measurement outcomes $|0\rangle$ and $|1\rangle$; the bar height marks 
the probability-weighted average (expectation value) of the residual 
coherence over the two possible measurement outcomes, and represents 
the expected coherence remaining after a random single-qubit measurement.
Dashed horizontal lines mark the initial coherence of each three-qubit
state before any measurement: $\mathcal{C}_{l_1}^{\mathrm{init}}\approx 1.914$
for $|\overline{W_3^L}\rangle$ (blue) and $\mathcal{C}_{l_1}^{\mathrm{init}}=2$
for $|W\rangle$ (red).
For the symmetric $|W\rangle$ state (red bars), the two possible
residual coherences are identical across all three qubits
($\mathcal{C}_{l_1}=1$ for outcome $|0\rangle$, $\mathcal{C}_{l_1}=0$ for outcome
$|1\rangle$), confirming the equilateral entanglement geometry.
For the asymmetric $|\overline{W_3^L}\rangle$ state (blue bars),
measuring the vertex qubit $A$ yields outcomes
$\mathcal{C}_{l_1}\in\{0,\,1\}$ with equal probability $\frac{1}{2}$, while 
measuring either base qubit $B$ or $C$ yields 
$\mathcal{C}_{l_1}\in\{0,\,\frac{2\sqrt{2}}{3}\approx 0.943\}$ with 
probabilities $\frac{1}{4}$ and $\frac{3}{4}$ respectively.
The higher bar height for the base qubits ($0.71$ vs $0.50$) 
quantitatively reflects the isosceles geometry: base qubits are 
biased toward the Hopf mode with probability $\frac{3}{4}$ (compared to 
$\frac{1}{2}$ for the vertex), and thus on average leave behind a more 
coherent residual state, even though the per-outcome coherence of 
the Hopf residual ($\approx 0.943$) is lower than that left by the 
vertex qubit ($1.0$).}
\label{fig:coherence_comparison}
\end{figure}

\subsection{Concurrence}
\label{subsec:concurrence}

For a two-qubit mixed state $\rho$, the concurrence is~\cite{Wootters1998}
\begin{eqnarray}
\label{concur1}
\mathcal{C}(\rho)=\max~\lbrace 0,\sqrt{\lambda_1}-\sqrt{\lambda_2}
-\sqrt{\lambda_3}-\sqrt{\lambda_4}\rbrace,
\end{eqnarray}
where $\lambda_i$'s are the eigenvalues of $\rho~\tilde{\rho}$ (in decreasing order). The spin-flipped density matrix $\tilde{\rho}$ is defined as
\begin{eqnarray}
\label{rhotilde}
    \tilde{\rho} = (\sigma_y\otimes\sigma_y)\rho^*(\sigma_y\otimes\sigma_y).
\end{eqnarray}

Here $\sigma_{y}$ is the Pauli spin matrix in the $y$-basis, $\tilde{\rho}$ is in the same basis as $\rho$, and  $\rho^{*}$ is the complex conjugate of the density matrix $\rho$.

All bipartite reduced states arising here are X-states 
(see Appendix~\ref{app:A}), for which \cite{rau2009algebraic,quesada2012quantum,mendoncca2014entanglement}
\begin{eqnarray}
\label{concur2}
\mathcal{C}(\rho) = 2\max \lbrace 0,|\rho_{23}|-\sqrt{\rho_{11}\rho_{44}}\rbrace,
\label{eq:C_Xstate}
\end{eqnarray}
where, $\rho_{ij}$'s denote the elements of the density matrix $\rho$.

Now, the symmetric $|W\rangle$ state distributes entanglement uniformly,
yielding a perfectly equilateral entanglement geometry with
$\mathcal{C}_W \approx 0.667$ for every bipartite reduction (here $\mathcal{C}_W$ is the concurrence). However, for
the asymmetric $|\overline{W_3^L}\rangle$ state, the situation is different. The entanglement as resource is concentrated as:
the vertex qubit acts as a central hub (Hopf 50\%, Borromean 
50\%), and a measurement outcome of $|1\rangle$ on the vertex 
leaves the remaining two qubits in a maximally entangled 
Bell state,
while base qubits $B$ and $C$ are biased switches (Hopf 75\%, Borromean 25\%, with
sub-maximal residual coherence $C_{l_1}\approx0.943$).
This structural asymmetry, together with the two-excitation sector,
forms the foundation for the divergent decoherence dynamics below.

\subsection{Noise Channels}
\label{subsec:channels}

We consider four single-qubit noise channels applied locally to one
qubit of each bipartite subsystem, leaving the other qubit isolated.
For the asymmetric $|\overline{W_3^L}\rangle$ state, the noise is
applied exclusively to the peripheral base qubits ($B$ or $C$), leaving
the vertex $A$ unperturbed.
Mathematically, this ensures a consistent basis of comparison between
$\mathcal{C}_{VB}$ and $\mathcal{C}_{BB}$; physically, it models a hub-and-spoke quantum network (i.e., a topology in which a single central node connects to multiple peripheral nodes) in which a protected central node communicates through noisy
peripheral channels~\cite{Kimble2008}. The four channels considered here are (a) \textit{phase damping} (pure dephasing,
parameter $p$), (b) \textit{amplitude damping} (energy dissipation,
parameter $\gamma$)~\cite{Plenio1998}, (c) \textit{depolarization}
(isotropic noise, parameter $p$)~\cite{Nielsen2010}, and (d) the
\textit{generalized amplitude damping channel} (GADC, parameters $p$
and $\alpha$)~\cite{Srikanth2008}.

Their Kraus operators, trace-preservation verification, and the
resulting decohered density-matrix elements are given in
Appendix~\ref{app:A}~\cite{Bennett1996}.

\section{Entanglement-Network Geometry}
\label{sec:geometry}

Before introducing environmental noise, we establish the structural baseline. The strength of the undecohered bipartite concurrence is dictated by how the Hopf and Borromean modes are shared among the qubits, together with the excitation number.

\subsection{The Equilateral Network of the Symmetric State}

The symmetric $|W\rangle$ state distributes its single excitation
uniformly~\cite{Dur2000}.
Every qubit triggers the Hopf mode with probability $\frac{2}{3}$ and the
Borromean mode with probability $\frac{1}{3}$; the Hopf outcome leaves a
maximally entangled Bell-like residual state (where coherence $\mathcal{C}_{l_1}=1$).
Tracing out any qubit yields the same reduced state, so all bipartite
concurrences are equal.

\begin{figure}[t!]
\centering
\includegraphics[width=0.8\columnwidth]{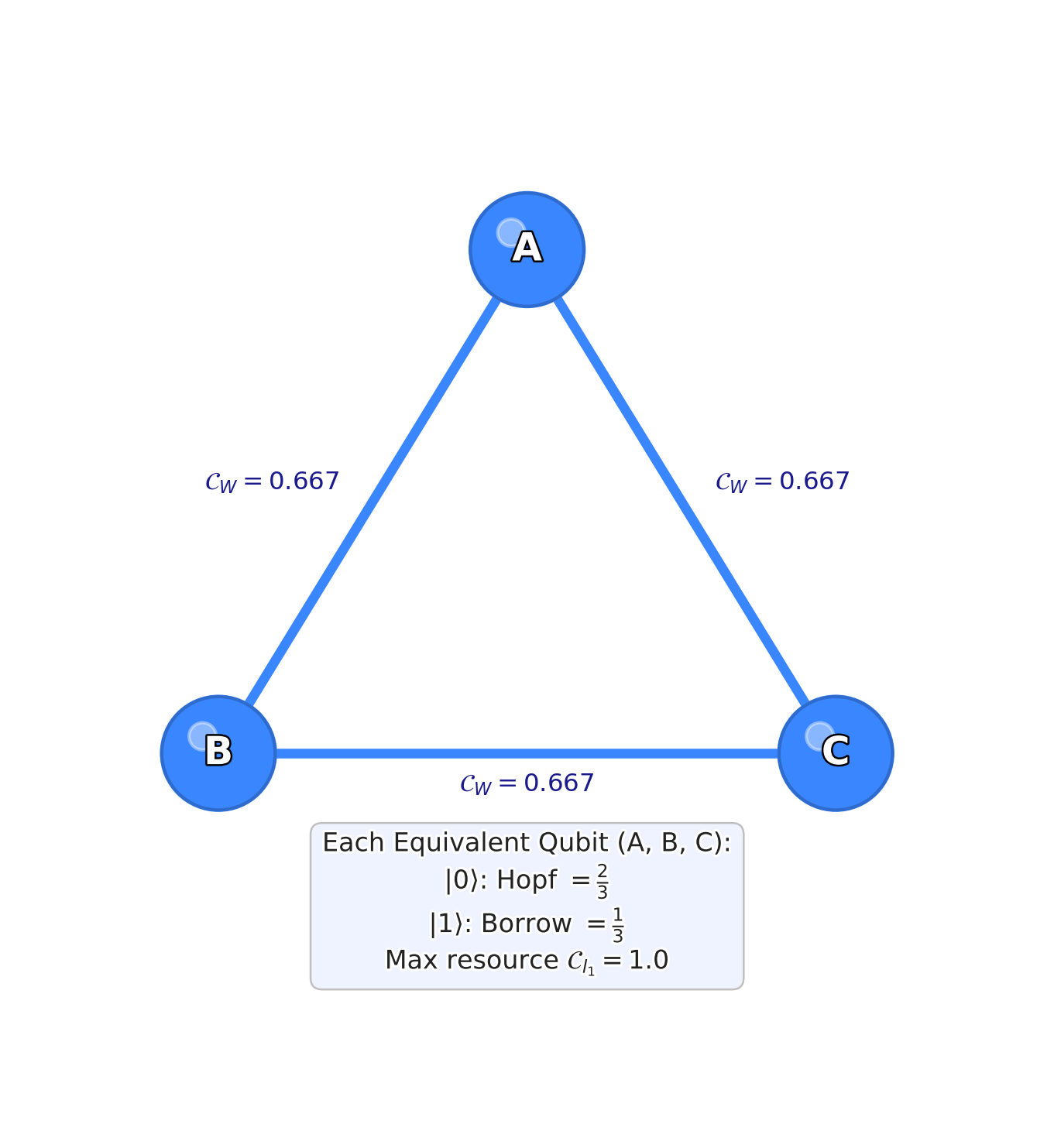}
\caption{\textbf{Equilateral entanglement-network geometry of the symmetric $|W\rangle$ state.} Each vertex represents a qubit ($A$, $B$, $C$) and each edge represents a bipartite entanglement link. Because the single excitation is distributed uniformly, all three qubits play structurally identical roles. Every qubit acts as an unbiased probabilistic switch, triggering the Hopf mode (residual coherence $\mathcal{C}_{l_1}=1.0$) with probability $\frac{2}{3}$ and the Borromean mode (residual separability, $\mathcal{C}_{l_1}=0$) with probability $\frac{1}{3}$. Consequently, tracing out any single qubit yields the same reduced density matrix, and all three bipartite concurrences are equal: 
$\mathcal{C}_W = \frac{2}{3} \approx 0.667$ 
(see Appendix~B, Eq.~(B2)). The geometry is perfectly equilateral, with no preferred hub or peripheral qubit.}
\label{fig:sym_topology}
\end{figure}
%\end{widetext}

%\begin{widetext}
\begin{figure}[t!]
\centering
\includegraphics[width=0.8\columnwidth]{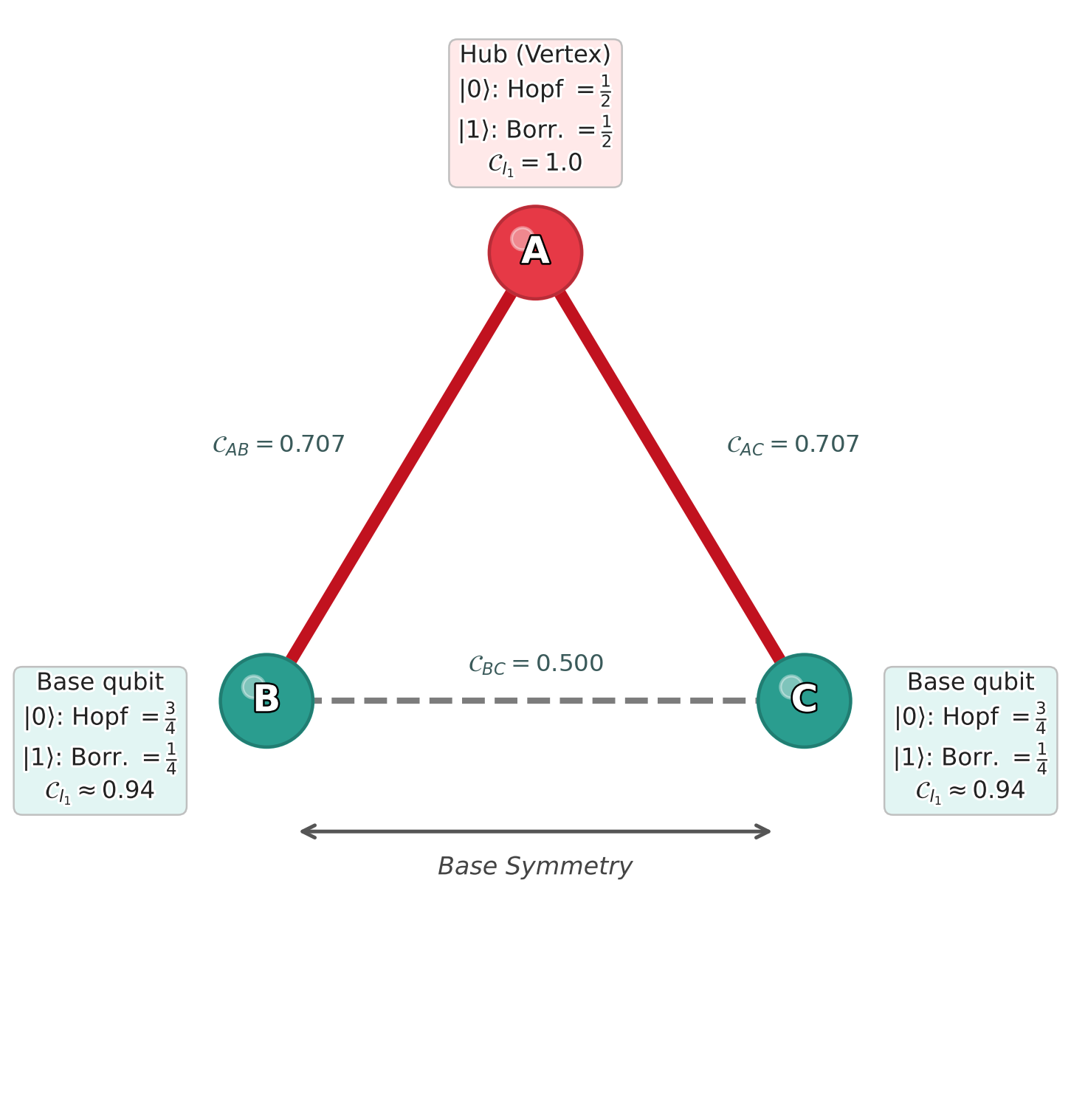}
\caption{\textbf{Isosceles entanglement-network geometry of the asymmetric $|\overline{W_3^L}\rangle$ state.} The vertex qubit $A$ (top) acts as a \textit{balanced switch}: it triggers the Hopf mode with probability $\frac{1}{2}$ and the Borromean mode with probability $\frac{1}{2}$, and its Hopf outcome leaves behind a maximally entangled Bell state (coherence $ \mathcal{C}_{l_1}=1.0$). The two base qubits $B$ and $C$ (bottom) each act as \textit{biased switches}: they trigger the Hopf mode more frequently (with probability $\frac{3}{4}$), but their Hopf outcome leaves a sub-maximally entangled residual state (coherence $\mathcal{C}_{l_1}\approx 0.943$). This asymmetry in the quality of the Hopf resource is reflected in the bipartite concurrences: the two vertex-base ($AB~ \&~ AC$) links form a stronger \textit{super-link} ($\mathcal{C}_{VB}=\frac{1}{\sqrt{2}} \approx 0.707$, solid edges), while the base-base ($BC$) link is weaker ($\mathcal{C}_{BB}=0.500$, dashed edge). The geometry is isosceles rather than equilateral, with $A$ as the privileged hub.}
\label{fig:asym_topology}
\end{figure}

\subsection{The Isosceles Network and the Super-Link}

The $|\overline{W_3^L}\rangle$ state (two-excitation subspace) breaks this equilateral symmetry~\cite{Lohmayer2006}, forming an isosceles geometry with vertex $A$ and base qubits $B$, $C$.

The initial concurrence hierarchy (derived in Appendix~\ref{app:B}) is:

\begin{align}
\label{initineq}
\mathcal{C}_{VB} = \frac{1}{\sqrt{2}} \approx 0.707 > 
\mathcal{C}_W &= \frac{2}{3} \approx 0.667 \nonumber\\
&> \mathcal{C}_{BB} = \frac{1}{2} = 0.500.
\end{align}

The results are summarized in table I.

\begin{table}[!b]
\centering
\caption{Baseline bipartite concurrences for the two states.}
\label{tab:baseline}
\resizebox{\columnwidth}{!}{%
\begin{threeparttable}
\begin{tabular}{llc}
\toprule
\textbf{State / Pair} & \textbf{Geometry} & 
\textbf{Initial Concurrence} $\mathcal{C}(0)$\tnote{$\dagger$} \\
\midrule
$|\overline{W_3^L}\rangle$ (vertex-base) & 
Isosceles super-link & 
$\dfrac{1}{\sqrt{2}} \approx 0.707$ \\[8pt]
$|W\rangle$ (any pair) & 
Equilateral & 
$\dfrac{2}{3} \approx 0.667$ \\[8pt]
$|\overline{W_3^L}\rangle$ (base-base) & 
Isosceles peripheral & 
$0.500$ \\
\bottomrule
\end{tabular}
\begin{tablenotes}
\footnotesize
\item[$\dagger$] $\mathcal{C}(0)$ denotes the bipartite 
concurrence evaluated at zero noise strength, i.e., for 
the undecohered state. Derivations are given in 
Appendix~B, Eqs.~(B2), (B4), and~(B6).
\end{tablenotes}
\end{threeparttable}%
}
\end{table}

This hierarchy raises the following key question: \textit{Does concentrating entanglement
into a super-link protect it from decoherence, or introduce additional
fragility?} We now answer this question by studying dynamics of each channel acting upon the state.

\section{Decoherence Dynamics}
\label{sec:decoherence}

We systematically apply four quantum noise channels locally to the bipartite networks. Analytical expressions are derived in Appendix~\ref{app:C}.

\subsection{Phase Damping: Preservation of the Initial Hierarchy}
\label{subsec:phase_damping}

Phase damping suppresses off-diagonal coherences without altering
populations~\cite{Nielsen2010}. Under local phase damping with scattering probability $p$,
$\rho_{23}\to\sqrt{1-p}\,\rho_{23}$ while $\rho_{11}$, $\rho_{44}$
are unchanged.
Hence all concurrences scale by the common factor $\sqrt{1-p}$. Thus we have
\begin{align}
\mathcal{C}_W(p) &= \frac{2}{3}\sqrt{1-p}, \\
\mathcal{C}_{VB}(p) &= \frac{1}{\sqrt{2}}\sqrt{1-p},\\
\mathcal{C}_{BB}(p) &= \frac{1}{2}\sqrt{1-p}.
\end{align}

\begin{figure}[!t]
\centering
\includegraphics[width=\columnwidth]{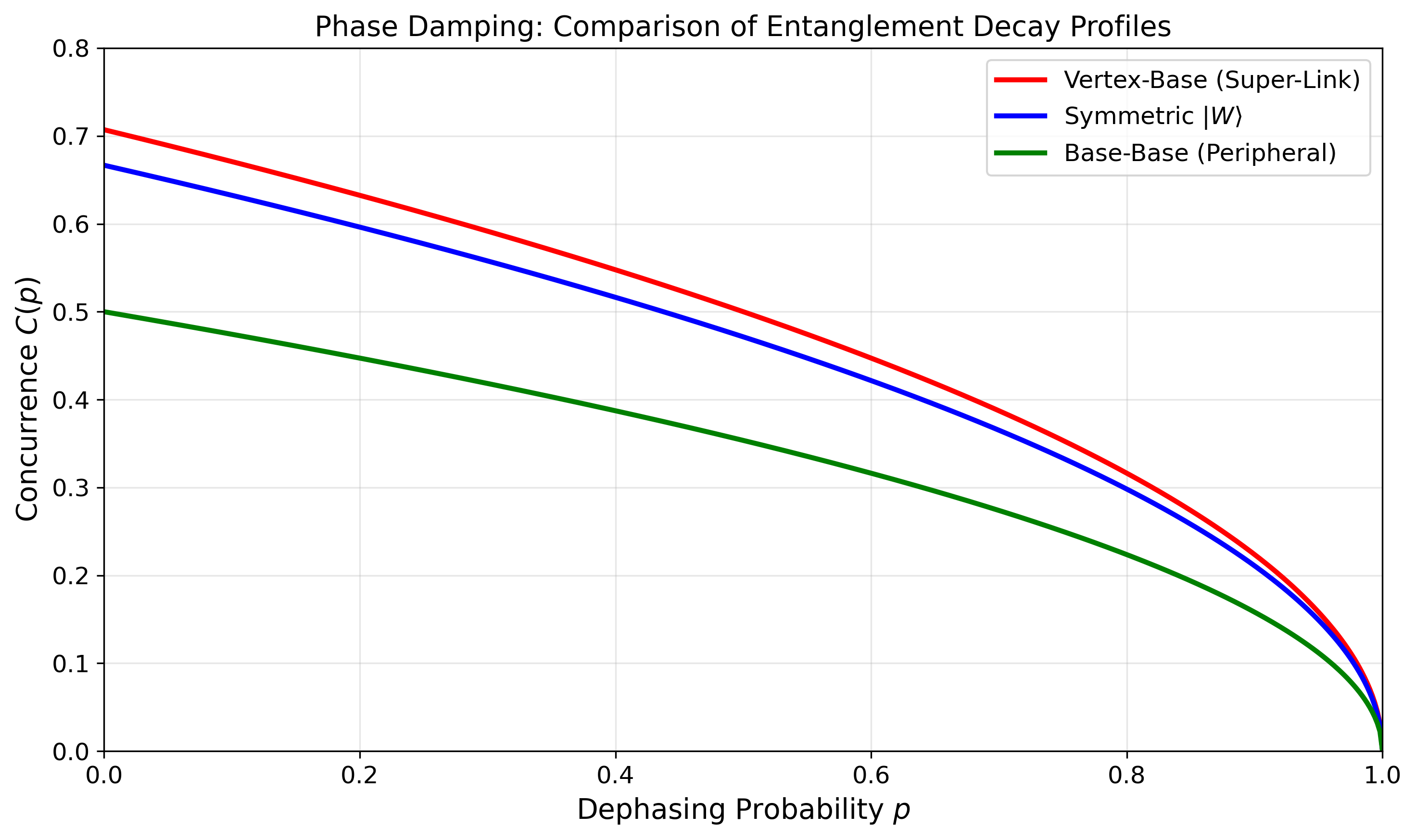}
\caption{\textbf{Bipartite concurrence decay under local phase damping.} All three concurrences — vertex-base super-link $\mathcal{C}_{VB}$ (red), symmetric $|W\rangle$ state $\mathcal{C}_{W}$ (blue), and base-base peripheral link $\mathcal{C}_{BB}$ (green) — decay smoothly and monotonically, each proportional to 
$\sqrt{1-p}$, where $p\in[0,1]$ is the dephasing probability. Because phase damping suppresses only off-diagonal coherences while leaving diagonal populations unchanged, it scales all concurrences by the same factor, preserving the initial hierarchy $\mathcal{C}_{VB} > \mathcal{C}_{W} > \mathcal{C}_{BB}$ for all $ p < 1 $. No entanglement sudden death (ESD) occurs: all three concurrences vanish only asymptotically at $p=1$. The super-link's advantage over the symmetric state is fully maintained under pure dephasing noise.}
\label{fig:phase_damping_comparison}
\end{figure}
The isosceles super-link retains its initial advantage that \textit{its higher
starting concurrence is preserved uniformly for all $p$}.

\subsection{Amplitude Damping: The Super-Link Fragility Effect}
\label{subsec:amplitude_damping}

Amplitude damping models energy dissipation to a zero-temperature
environment with decay probability $\gamma$~\cite{Plenio1998}.
The concurrence dynamics are:
\begin{align}
\mathcal{C}_W(\gamma) &= \frac{2}{3}\sqrt{1-\gamma}, \label{eq:CW_AD}\\
\mathcal{C}_{VB}(\gamma) &= \frac{\sqrt{1-\gamma}}{\sqrt{2}}\bigl(1-\sqrt{\gamma}\bigr),
\label{eq:CVB_AD}\\
\mathcal{C}_{BB}(\gamma) &= \frac{\sqrt{1-\gamma}}{2}
\bigl(1-\sqrt{2\gamma}\bigr)_{+},
\label{eq:CBB_AD}
\end{align}
where$(x)_+=\max~\lbrace 0,x\rbrace $.

\begin{figure}[!t]
\centering
\includegraphics[width=\columnwidth]{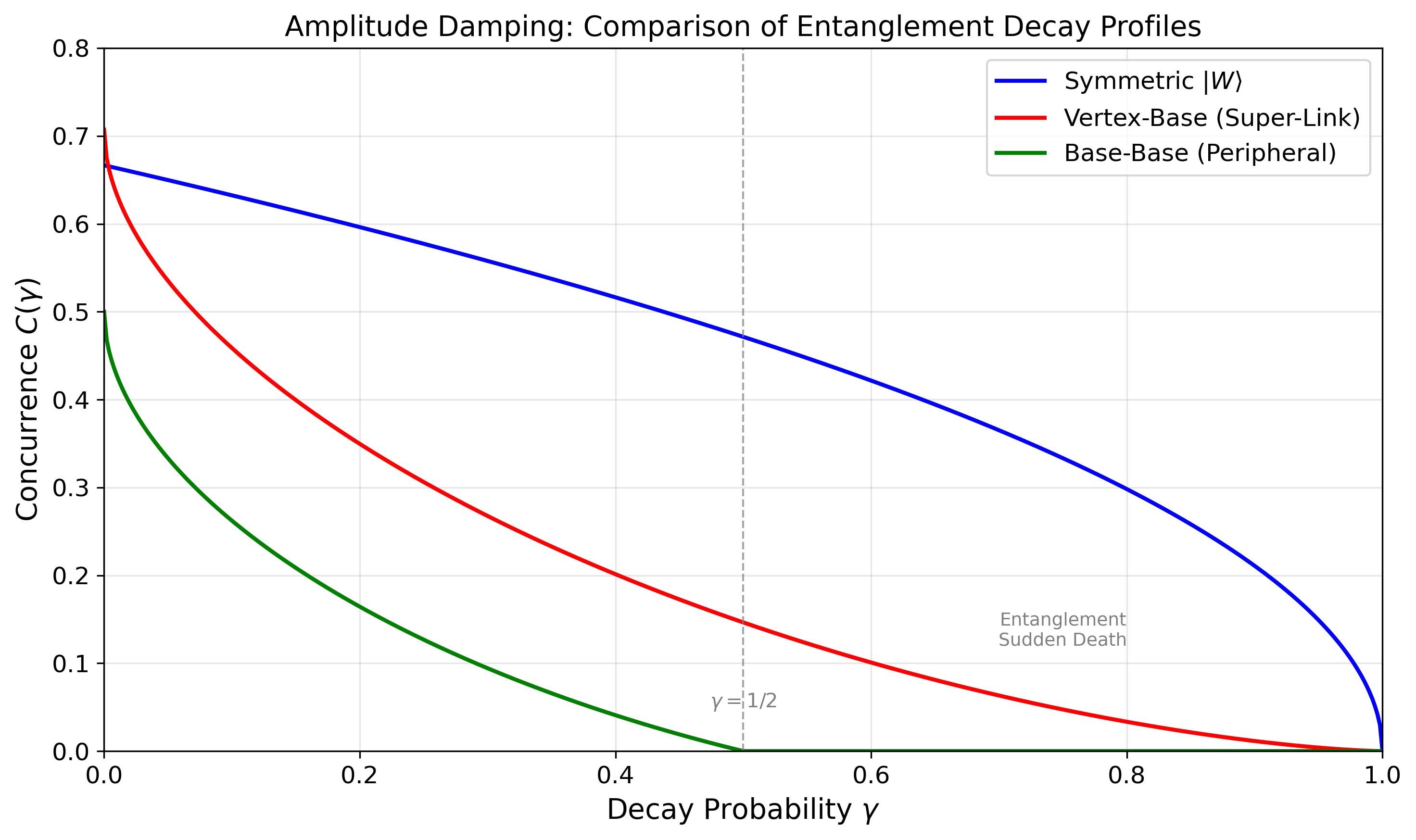}
\caption{\textbf{Bipartite concurrence decay under local amplitude
damping, illustrating the Super-Link Fragility Effect.}
As the decay probability $\gamma$ increases from $0$ to $1$, the
initial hierarchy $\mathcal{C}_{VB} > \mathcal{C}_W > \mathcal{C}_{BB}$ is reordered.
The symmetric $|W\rangle$ state (blue) is the most robust, decaying
only as $\frac{2}{3}\sqrt{1-\gamma}$ with no ESD.
The vertex-base super-link $\mathcal{C}_{VB}$ (red) decays faster due to an
additional factor $(1-\sqrt{\gamma})$ arising from the double-excitation
terms $|110\rangle$, $|101\rangle$ of $|\overline{W_3^L}\rangle$.
The base-base link $\mathcal{C}_{BB}$ (green) undergoes entanglement sudden death
at $\gamma=\frac{1}{2}$ (vertical dashed line), becoming identically zero for
all $\gamma\geq \frac{1}{2}$.}
\label{fig:amplitude_damping_comparison}
\end{figure}

The reversal is driven by two concurring factors.
First, the $|\overline{W_3^L}\rangle$ state lies in the two-excitation
subspace; its multi-excitation terms ($|110\rangle$, $|101\rangle$) are
heavily penalised by amplitude damping, which drains excited-state
populations toward $|000\rangle$.This means, under amplitude
damping, components of the quantum state associated with $|110\rangle$, $|101\rangle$ lose probability mass rapidly and their amplitudes decrease faster than states with fewer excitations, eventually population accumulates near the ground state $|000\rangle$.
Second, the structural asymmetry concentrates entanglement in the
super-link, but this very concentration makes it sensitive to
population redistribution.
The single-excitation $|W\rangle$ state, containing no double-excitation
terms, suffers only asymptotic decay. These two factors together explain the named effect summarized below.

\textbf{Super-Link Fragility Effect:} concentrating bipartite
entanglement into a stronger link increases its susceptibility to
energy-dissipative noise when that link is supported by multi-excitation
amplitudes.
We note that, while the hierarchy reversal is partly expected from the
excitation-sector difference alone, the quantitative comparison of
exact ESD thresholds and decay rates requires the full analytical
treatment given here.

\subsection{Depolarization: Erasing the Asymmetry Advantage}
\label{subsec:depolarization}

Depolarization isotropically scrambles the qubit state with probability
$p$~\cite{Nielsen2010}.
For $p \in [0,\frac{3}{4}]$ the concurrence expressions are:
\begin{align}
\mathcal{C}_W(p) &= \max\! \Big\lbrace 0,\frac{2}{9}(3-4p-\sqrt{6p})\Big\rbrace,
\label{eq:CW_dep}\\
\mathcal{C}_{VB}(p) &= \max\! \Big\lbrace 0,\frac{\sqrt{2}}{6}(3-4p-\sqrt{6p})\Big\rbrace,
\label{eq:CVB_dep}\\
\mathcal{C}_{BB}(p) &= \max\! \Big\lbrace 0,\frac{1}{6}
\bigl(3-4p-2\sqrt{3p-p^2}\bigr)\Big\rbrace.
\label{eq:CBB_dep}
\end{align}
Since $\mathcal{C}_{VB}(p)$ and $\mathcal{C}_W(p)$ differ 
only by the constant multiplicative factor $\frac{3\sqrt{2}}{4}$ 
(see Eqs.~\eqref{eq:CW_dep}--\eqref{eq:CVB_dep}), they vanish 
at the same threshold $p = 3/8 = 0.375$. The base-base link 
$\mathcal{C}_{BB}$, governed by a structurally different 
expression, reaches zero earlier at $p = 3/10 = 0.300$.

\begin{figure}[!t]
\centering
\includegraphics[width=\columnwidth]{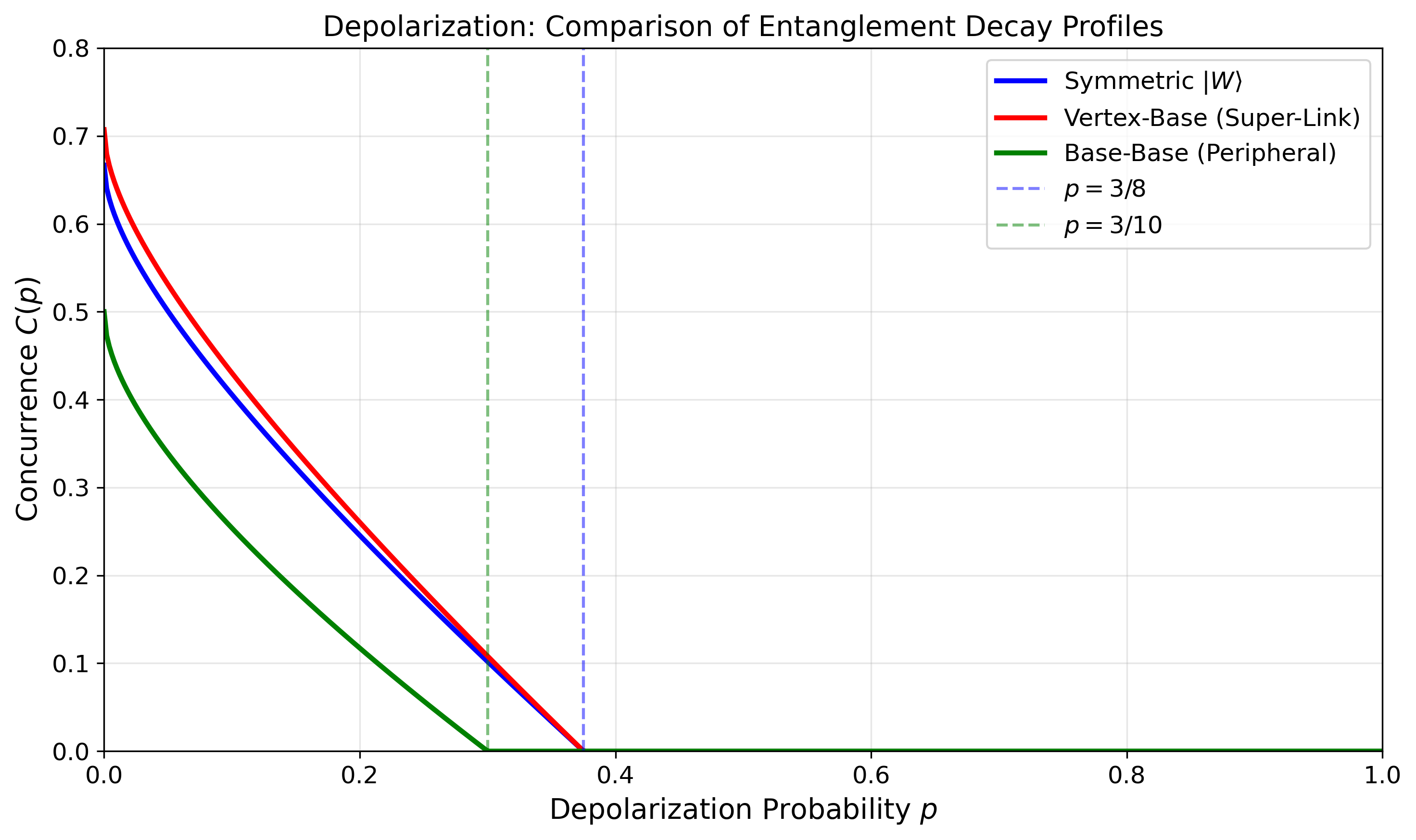}
\caption{\textbf{Bipartite concurrence decay under local depolarization.}
The depolarization probability $p\in[0,1]$ parametrizes isotropic
qubit scrambling.
The base-base link $\mathcal{C}_{BB}$ (green) undergoes entanglement sudden death (ESD)
first, at $p=\frac{3}{10} =0.300$ (right vertical dashed line).
Both $\mathcal{C}_W$ (blue) and $\mathcal{C}_{VB}$ (red) survive longer but undergo
entanglement sudden death simultaneously at $p=\frac{3}{8}=0.375$ (left
vertical dashed line), despite their different initial values.
This coincidence of ESD thresholds shows that the geometric asymmetry
advantage of the super-link is completely erased by isotropic noise:
the additional initial concurrence of $\mathcal{C}_{VB}$ does not translate into
any survival benefit. Under depolarization, noise symmetry overwhelms
state asymmetry.}
\label{fig:depolarization_comparison}
\end{figure}

Despite its higher initial concurrence, the super-link offers no
survival advantage under isotropic noise~\cite{Huang2007}.
This can be understood directly from 
Eqs.~\eqref{eq:CW_dep}--\eqref{eq:CVB_dep}: 
$\mathcal{C}_{VB}$ and $\mathcal{C}_W$ are proportional 
by the constant factor obtained by dividing 
Eq.~\eqref{eq:CVB_dep} by Eq.~\eqref{eq:CW_dep}:

\begin{equation}
\frac{\mathcal{C}_{VB}(p)}{\mathcal{C}_W(p)} = 
\frac{\dfrac{\sqrt{2}}{6}\left(3 - 4p - \sqrt{6p}\right)}
{\dfrac{2}{9}\left(3 - 4p - \sqrt{6p}\right)}  = \frac{3\sqrt{2}}{4} \approx 1.061,
\end{equation}

and therefore they share the same zero, i.e., they vanish 
at the same threshold  at $p = \frac{3}{8} = 0.375$.

\subsection{Generalized Amplitude Damping: Interpolation Between
Regimes}
\label{subsec:gadc}

The GADC models a finite-temperature reservoir with interaction strength
$p$ and a parameter $\alpha \in [0,1]$ weighting damping ($\alpha$)
against excitation ($1-\alpha$)~\cite{Srikanth2008}.
%% [Comment 14: alpha=0 limit explained carefully]
Here $\alpha=1$ recovers pure amplitude damping ($T=0$), while
$\alpha=0$ corresponds to the pure excitation limit of the channel
(the environment only pumps the qubit into $|1\rangle$; this is a
mathematical limit of the GADC parametrization and should not be
over-interpreted as a physical high-temperature regime without
specifying the full thermal model).
The concurrence expressions are:
\begin{align}
\mathcal{C}_W(p,\alpha) &= \max\!\Big\lbrace 0,\tfrac{2}{3}\bigl[\sqrt{1-p}\notag\\
&\quad -\sqrt{(1-p+2\alpha p)(1-\alpha)p}\,\bigr]\Big\rbrace,
\label{eq:CW_GADC}\\
\mathcal{C}_{VB}(p,\alpha) &= \max\!\Big\lbrace 0,\tfrac{1}{\sqrt{2}}\bigl[\sqrt{1-p}\notag\\
&\quad -\sqrt{\alpha p(1+p-2\alpha p)}\,\bigr]\Big\rbrace,
\label{eq:CVB_GADC}\\
\mathcal{C}_{BB}(p,\alpha) &= \max\!\Big\lbrace 0,\tfrac{1}{2}\bigl[\sqrt{1-p}\notag\\
&\quad -\sqrt{\alpha p(2+p-3\alpha p)}\,\bigr]\Big\rbrace.
\label{eq:CBB_GADC}
\end{align}
At $\alpha=0$, the second terms under each max vanish and all
concurrences reduce to the phase-damping-like form proportional to $\sqrt{1-p}$, restoring the original hierarchy.

\begin{figure*}[!ht]
\centering
\includegraphics[width=0.95\textwidth]{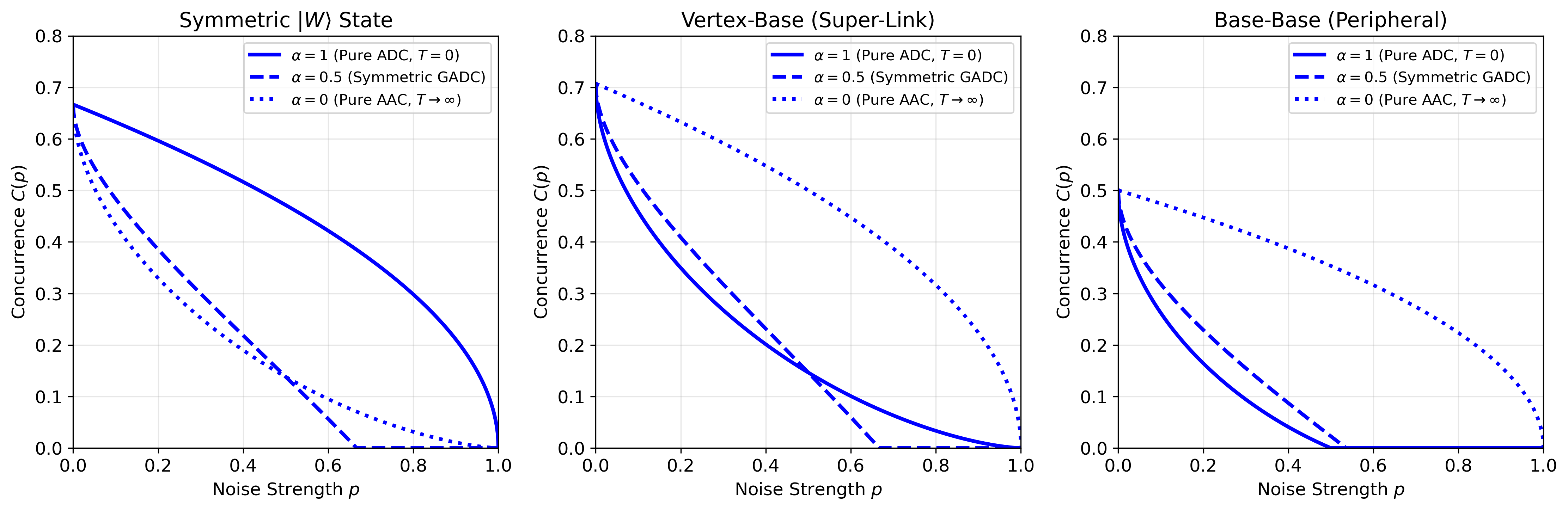}
\caption{\textbf{Bipartite concurrence decay under the GADC for all
three bipartite links, each shown across three values of the
damping-excitation weight $\alpha$.}
In every panel, the solid line corresponds to $\alpha=1$ (pure
amplitude damping, $T=0$), the dashed line to $\alpha=0.5$ (symmetric
exchange), and the dotted line to $\alpha=0$ (pure excitation limit of
the GADC).
\textit{Left} (symmetric $|W\rangle$): the state is most robust at
$\alpha=1$ and decays more rapidly as $\alpha$ decreases toward 0,
reflecting sensitivity of the single-excitation state to excitation
noise.
\textit{Centre} (vertex-base super-link $\mathcal{C}_{VB}$): the super-link is
most fragile at $\alpha=1$ but becomes increasingly robust as
$\alpha\to 0$, directly visualising the channel-dependence of the
Super-Link Fragility Effect.
\textit{Right} (base-base link $\mathcal{C}_{BB}$): robustness improves as
$\alpha$ decreases, and ESD is progressively delayed.
The GADC interpolates continuously between the hierarchy-reversing
amplitude-damping limit ($\alpha=1$) and the
hierarchy-preserving excitation limit ($\alpha=0$).}
\label{fig:gad_comparison_all_bonds}
\end{figure*}

\begin{figure*}[!ht]
\centering
\begin{subfigure}{0.32\textwidth}
\centering
\includegraphics[width=\linewidth]{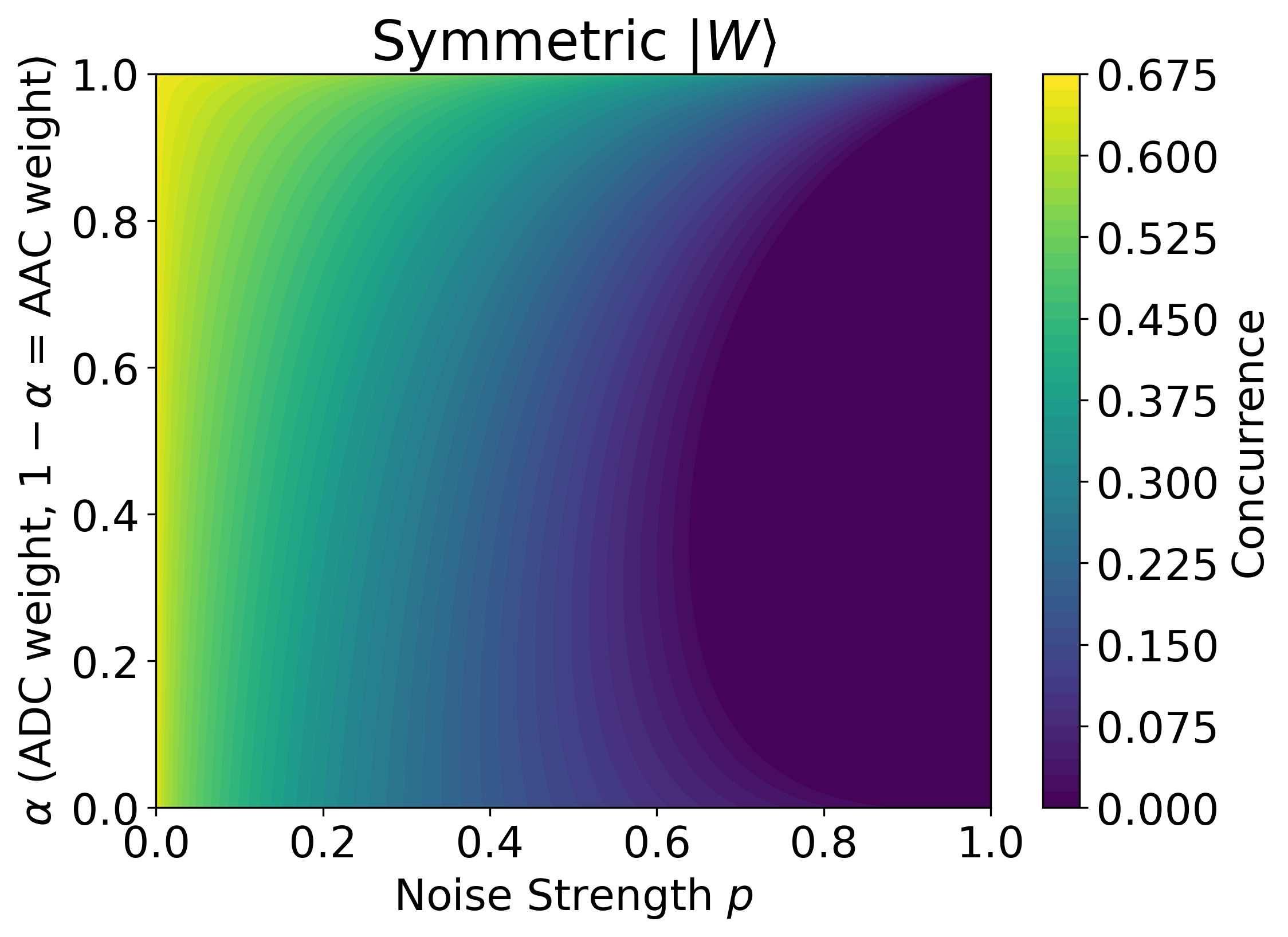}
%\caption{Symmetric $|W\rangle$ state}
\label{fig:gad_heatmap_sym}
\end{subfigure}
\hfill
\begin{subfigure}{0.32\textwidth}
\centering
\includegraphics[width=\linewidth]{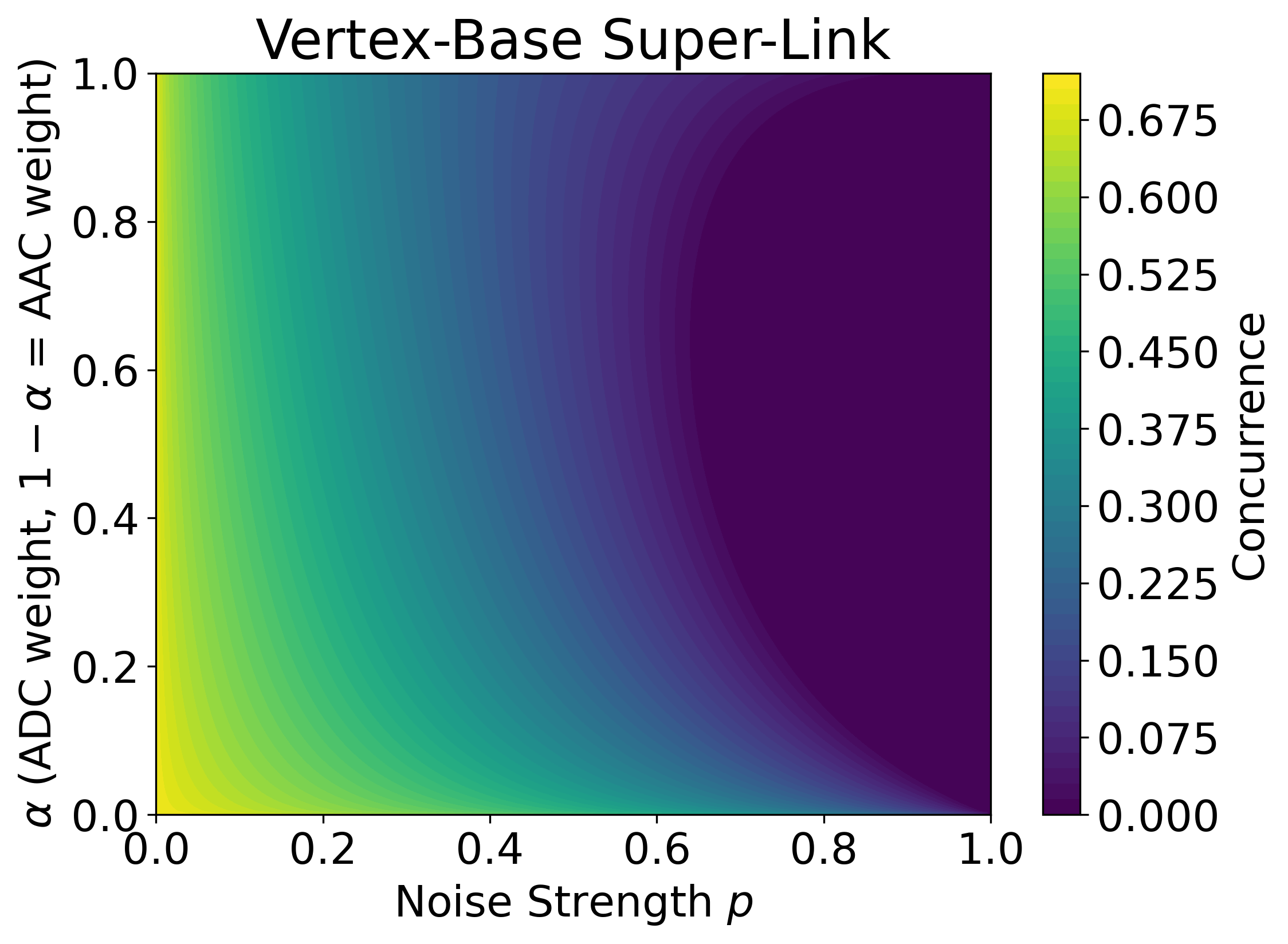}
%\caption{Vertex-base super-link}
\label{fig:gad_heatmap_vb}
\end{subfigure}
\hfill
\begin{subfigure}{0.32\textwidth}
\centering
\includegraphics[width=\linewidth]{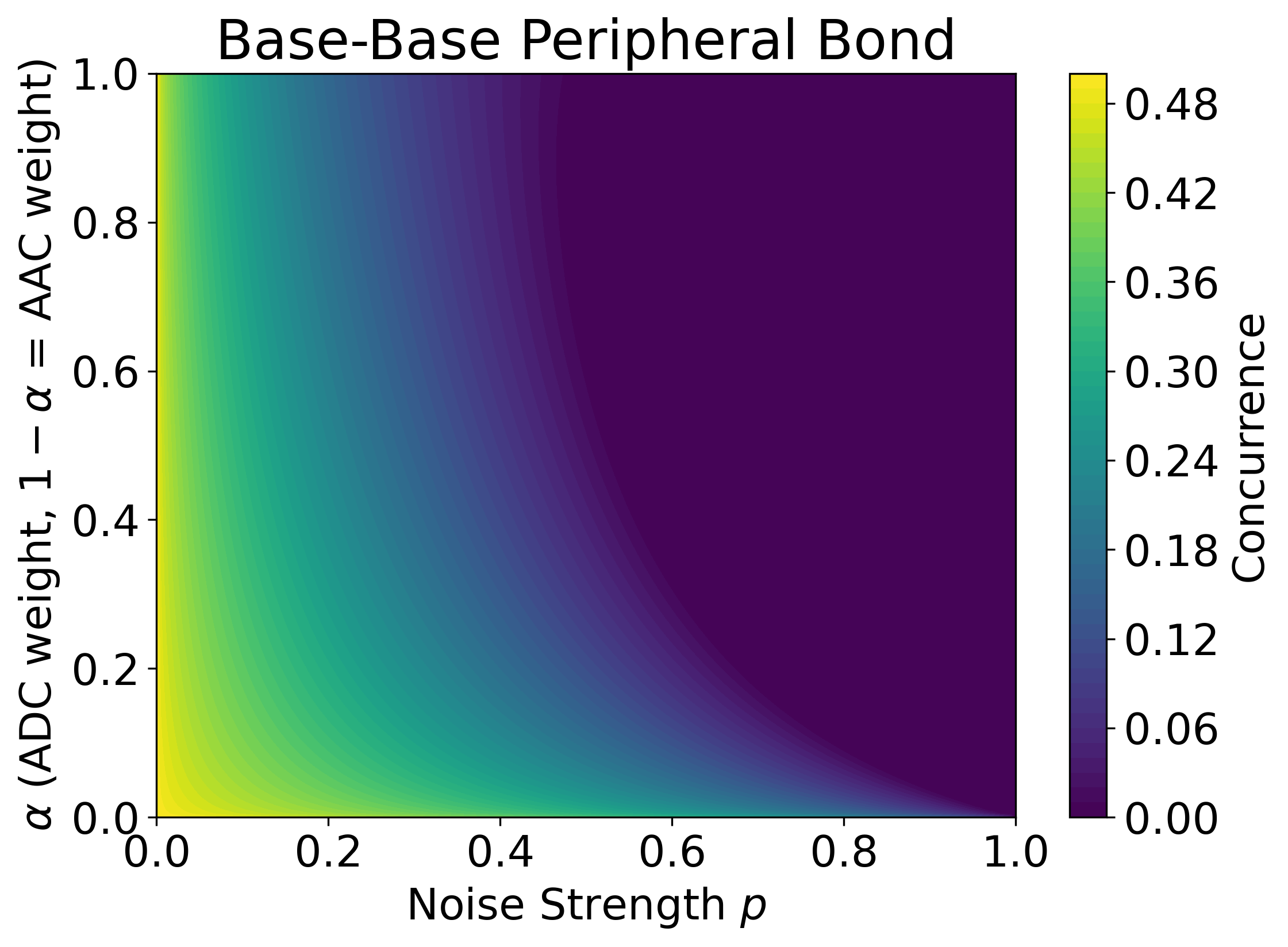}
%\caption{Base-base link}
\label{fig:gad_heatmap_bb}
\end{subfigure}
\caption{\textbf{Concurrence heatmaps in the $(p,\,\alpha)$ plane under the GADC.} Each panel shows the bipartite concurrence as a function of noise strength $p\in[0,1]$ (horizontal axis) and damping weight $\alpha\in[0,1]$ (vertical axis), where $\alpha=1$ is pure amplitude damping and $\alpha=0$ is pure excitation. Lighter (warmer) colours indicate higher concurrence; dark regions correspond to zero concurrence (separable state). \textit{(a)} The symmetric $|W\rangle$ state maintains a broad entangled region across the entire $\alpha$ range due to its single-excitation structure. \textit{(b)} The vertex-base super-link is fragile near $\alpha=1$ (narrow entangled region) but robust near $\alpha=0$ (broad region), directly visualising the channel-dependence of the Super-Link Fragility Effect. \textit{(c)} The base-base peripheral bond has the most restricted entangled region at all $\alpha$, consistent with its lowest initial concurrence.}
\label{fig:gad_heatmaps}
\end{figure*}

\begin{table*}[b]
\centering
\caption{Channel-dependent entanglement hierarchy and ESD thresholds.}
\label{tab:hierarchy_summary}
\renewcommand{\arraystretch}{1.75}
\begin{tabular}{|l|c|l|l|}
\hline
\textbf{Channel} & \textbf{Mechanism} & \textbf{Hierarchy}
& \textbf{ESD thresholds} \\
\hline
Phase damping & Dephasing & $\mathcal{C}_{VB}>\mathcal{C}_W>\mathcal{C}_{BB}$ & None \\
Amplitude damping & Energy dissipation
& $\mathcal{C}_W > \mathcal{C}_{VB} > \mathcal{C}_{BB}$ & $\mathcal{C}_{BB}$: $\gamma=\frac{1}{2}$ \\
Depolarization & Isotropic scrambling
& $\mathcal{C}_{VB}\approx \mathcal{C}_W > \mathcal{C}_{BB}$
& $\mathcal{C}_{BB}$: $p=\frac{3}{10}$; $\mathcal{C}_W, \mathcal{C}_{VB}$: $p=\frac{3}{8}$ \\
GADC ($\alpha=0$) & Pure excitation (limit)
& $\mathcal{C}_{VB} > \mathcal{C}_W > \mathcal{C}_{BB}$ & None \\
\hline
\end{tabular}
\end{table*}

\section{Synthesis: A Unifying Framework}
\label{sec:synthesis}

Our results show that the initial concurrence hierarchy is not a
reliable predictor of decoherence robustness.
Instead, robustness emerges from the interplay of three factors:
\begin{enumerate}
    \item \textit{Entanglement-network geometry} determines the initial
    concurrences via the Hopf/Borromean mode distribution.
    \item \textit{Excitation-sector}: single-excitation states (symmetric
    $|W\rangle$) are resilient to amplitude damping; multi-excitation
    states ($|\overline{W_3^L}\rangle$) are vulnerable because their
    double-excitation amplitudes are strongly penalised by energy
    dissipation.
    \item \textit{Noise symmetry}: anisotropic channels (dephasing,
    amplitude damping) amplify or suppress the geometric asymmetry;
    isotropic channels (depolarization) wash it out.
\end{enumerate}

Table~\ref{tab:hierarchy_summary} summarises the channel-dependent
hierarchy.

\subsection{Qualitative Design Considerations for Quantum Networks}
\label{subsec:design}

Based on the analytical results above, we suggest the following
qualitative design considerations for Noisy Intermediate-Scale Quantum (NISQ)-era
architectures~\cite{Preskill2018,Coffman2000,Breuer2007,Zurek2003}.
These are indicative rather than prescriptive, since realistic hardware
noise models involve additional complexity beyond the idealized single-qubit
channels studied here.

\begin{itemize}
    \item \textsc{Dissipation-dominated platforms} (e.g.\ superconducting
    qubits at millikelvin temperatures): the symmetric $|W\rangle$ state,
    lying in the single-excitation subspace, is preferable for preserving
    bipartite entanglement links and avoiding ESD.

    \item \textsc{Dephasing-dominated platforms} (e.g.\ certain
    solid-state spin systems with long $T_1$ but short $T_2$): the
    asymmetric $|\overline{W_3^L}\rangle$ state offers a protected
    super-link, since under phase damping the hierarchy
    $\mathcal{C}_{VB} > \mathcal{C}_W > \mathcal{C}_{BB}$ is maintained for all $p < 1$.

    \item \textsc{Environments with significant thermal excitation}
    (the GADC regime with $\alpha\ll1$): the super-link advantage may
    be partially restored when excitation rates compete with damping,
    as shown in Section~\ref{subsec:gadc}.
\end{itemize}

\subsection{Future Directions}

Generalizing to $N$-qubit $W$-class states~\cite{Espoukeh2015},
incorporating non-Markovian dynamics~\cite{Breuer2007,Bellomo2007}, and
experimental verification in trapped-ion or superconducting
platforms~\cite{Preskill2018} are natural extensions of our work.
The interplay of entanglement and quantum coherence~\cite{Baumgratz2014}
in asymmetric states under finite-temperature reservoirs also warrants
further study~\cite{Ma2012,Ozaydin2015,Toth2012}.

\section{Conclusion}
\label{sec:conclusion}

We have analytically derived the concurrence dynamics of two
three-qubit $W$-class states; (a) the single-excitation symmetric
$|W\rangle$ and (b) the two-excitation asymmetric $|\overline{W_3^L}\rangle$, 
under four local quantum noise channels.
The central finding is that the initial bipartite entanglement hierarchy
$\mathcal{C}_{VB} > \mathcal{C}_W> \mathcal{C}_{BB}$ is not a reliable predictor of decoherence
robustness; depending on the channel, it can be preserved, erased, or
reversed. Specifically, under phase damping, all concurrences scale identically
by $\sqrt{1-p}$ and the hierarchy is preserved with no ESD.
Under amplitude damping, the hierarchy is reordered ($\mathcal{C}_W > \mathcal{C}_{VB} > \mathcal{C}_{BB}$),
with $\mathcal{C}_{BB}$ undergoing ESD at $\gamma= \frac{1}{2}$; this is the \textit{Super-Link
Fragility Effect}, driven jointly by structural asymmetry and the
two-excitation sector of $|\overline{W_3^L}\rangle$.
Under depolarization, the asymmetry advantage is erased: $\mathcal{C}_W$ and
$\mathcal{C}_{VB}$ share the ESD threshold $p=\frac{3}{8}$, while $\mathcal{C}_{BB}$ dies at
$p=\frac{3}{10}$. The GADC interpolates between these regimes continuously as $\alpha$
varies from $1$ (amplitude-damping limit) to $0$ (pure-excitation limit),
restoring the original hierarchy at $\alpha=0$. These results provide a systematic analytical framework linking
entanglement robustness to network geometry, excitation number, and
noise symmetry, and offer qualitative guidance for noise-resilient
quantum state selection in NISQ-era distributed quantum systems.

\section*{Acknowledgments}
F.O. acknowledges financial support from Tokyo International University Personal Research Fund and Special Grant-in-Aid for Research Work.

\addcontentsline{toc}{section}{Appendices}
\appendix
\section{Technical Notes and Analytical Methods}
\label{app:A}

This section establishes the mathematical toolkits utilized throughout the supplementary derivations, specifically focusing on the analytical evaluation of bipartite entanglement for X-states and the operator-sum representation of quantum noise.

\subsection{Derivation of the X-State Concurrence Formula}
\label{subsec:S1_xstate}

To quantify bipartite entanglement, we employ Wootters' concurrence \cite{Wootters1998,Hill1997,Mintert2005}. For a general two-qubit mixed state $\rho$, $\mathcal{C}(\rho) = \max\{0, \sqrt{\lambda_1} - \sqrt{\lambda_2} - \sqrt{\lambda_3} - \sqrt{\lambda_4}\}$, where $\lambda_i$s are the eigenvalues, in decreasing order, of the matrix $R = \rho \tilde{\rho}$, and $\tilde{\rho} = (\sigma_y \otimes \sigma_y) \rho^* (\sigma_y \otimes \sigma_y)$ is the spin-flipped state.

Throughout this work, the reduced density matrices obtained after tracing out a third qubit persistently take the form of X-states:
\begin{equation}
\rho_X = \begin{pmatrix}
\rho_{11} & 0 & 0 & \rho_{14} \\
0 & \rho_{22} & \rho_{23} & 0 \\
0 & \rho_{32} & \rho_{33} & 0 \\
\rho_{41} & 0 & 0 & \rho_{44}
\end{pmatrix}.
\end{equation}

For real density matrices ($\rho_{ij} = \rho_{ji}$), the spin-flipped matrix $\tilde{\rho}_X$ results in the product $R = \rho_X \tilde{\rho}_X$ having a strict block-diagonal structure. The eigenvalues of this product isolate into two independent $2 \times 2$ blocks.

The eigenvalues of these corner and center blocks are exactly:
\begin{align}
\lambda_{\pm}^{\text{corner}} &= \left(\sqrt{\rho_{11}\rho_{44}} \pm |\rho_{14}|\right)^2, \\
\lambda_{\pm}^{\text{center}} &= \left(\sqrt{\rho_{22}\rho_{33}} \pm |\rho_{23}|\right)^2.
\end{align}

Extracting the square roots of these eigenvalues and applying Wootters' formula, the concurrence simplifies to:
\begin{equation}
\mathcal{C}(\rho_X) = 2\max \lbrace 0, |\rho_{23}| - \sqrt{\rho_{11}\rho_{44}}, |\rho_{14}| - \sqrt{\rho_{22}\rho_{33}}\rbrace.
\end{equation}

Because every initial and decohered state evaluated in this study strictly maintains $\rho_{14} = \rho_{41} = 0$, the expression permanently reduces to our master analytical formula:
\begin{equation}
\boxed{\mathcal{C}(\rho_X) = 2\max \lbrace 0, |\rho_{23}| - \sqrt{\rho_{11}\rho_{44}} \rbrace.}
\label{eq:master_concurrence}
\end{equation}
Having rigorously established this reduction, Eq.~\ref{eq:master_concurrence} is used exclusively for all concurrence derivations in the subsequent sections.

\subsection{Trace Preservation Verification}
\label{subsec:S1_trace}

For each quantum channel evaluated, the applied single-qubit Kraus operators satisfy the completeness relation $\sum_k K_k^\dagger K_k = I$. Consequently, the two-qubit operators $E_k = I \otimes K_k$ naturally preserve the trace of the bipartite density matrices: $\operatorname{Tr}(\rho') = \operatorname{Tr}\left(\sum_k E_k \rho E_k^\dagger\right) = 1$. Trace conservation was explicitly verified at the conclusion of every matrix summation in Appendix~\ref{app:C}.

\subsection{Kraus Representations of Noise Channels}

\noindent \textbf{1. Phase Damping (Pure Dephasing):} For a scattering probability $p$, the Kraus operators are:
\begin{equation}
K_0 = \begin{pmatrix} 1 & 0 \\ 0 & \sqrt{1-p} \end{pmatrix}, \quad K_1 = \begin{pmatrix} 0 & 0 \\ 0 & \sqrt{p}. \end{pmatrix}
\end{equation}

\noindent \textbf{2. Amplitude Damping (Energy Dissipation):} This channel models dissipative energy transfer to a zero-temperature environment with decay probability $\gamma$ \cite{Plenio1998}:
\begin{equation}
K_0 = \begin{pmatrix} 1 & 0 \\ 0 & \sqrt{1-\gamma} \end{pmatrix}, \quad K_1 = \begin{pmatrix} 0 & \sqrt{\gamma} \\ 0 & 0 \end{pmatrix}.
\end{equation}

\noindent \textbf{3. Depolarization (Isotropic Noise):} Depolarization describes isotropic noise where the qubit has a probability $p$ of being replaced by the completely mixed state $\frac{I}{2}$, modelled using the Pauli matrices ($\sigma_x, \sigma_y, \sigma_z$) \cite{Nielsen2010}:

\begin{equation}
\begin{aligned}
K_0 = \sqrt{1-p}~ I, \quad & K_1 = \sqrt{\frac{p}{3}}~ \sigma_x, \\
K_2 = \sqrt{\frac{p}{3}}~ \sigma_y, \quad & K_3 = \sqrt{\frac{p}{3}}~ \sigma_z.
\end{aligned}
\end{equation}

\noindent \textbf{4. Generalized Amplitude Damping (GADC):} To model interactions with a finite-temperature thermal reservoir, the GADC introduces an interaction strength $p$ and temperature weights $\alpha$ (probability of transitioning to the ground state) and $\beta=1-\alpha$ (probability of transitioning to the excited state) \cite{Srikanth2008}:
\begin{equation}
\begin{aligned}
K_0 &= \sqrt{\alpha} \begin{pmatrix} 1 & 0 \\ 0 & \sqrt{1-p} \end{pmatrix}, \quad &K_1 &= \sqrt{\alpha p} \begin{pmatrix} 0 & 1 \\ 0 & 0 \end{pmatrix}, \\
K_2 &= \sqrt{\beta} \begin{pmatrix} \sqrt{1-p} & 0 \\ 0 & 1 \end{pmatrix}, \quad &K_3 &= \sqrt{\beta p} \begin{pmatrix} 0 & 0 \\ 1 & 0 \end{pmatrix}.
\end{aligned}
\end{equation}

\section{State Definitions and Initial Concurrence}
\label{app:B}

\subsection{Symmetric $|W\rangle$ State}

The symmetric W state is defined as $|W\rangle = \frac{1}{\sqrt{3}}(|001\rangle + |010\rangle + |100\rangle)$ \cite{Dur2000}. Tracing out any single qubit (e.g., qubit A) yields identical bipartite reduced density matrices:
\begin{equation}
\rho_{BC}^{(W)} = \operatorname{Tr}_A(|W\rangle\langle W|) = \begin{pmatrix}
\frac{1}{3} & 0 & 0 & 0 \\
0 & \frac{1}{3} & \frac{1}{3} & 0 \\
0 & \frac{1}{3} & \frac{1}{3} & 0 \\
0 & 0 & 0 & 0
\end{pmatrix}.
\end{equation}
Applying Eq.~\ref{eq:master_concurrence}:
\begin{equation}
\mathcal{C}_W(0) = 2\max\! \Big \lbrace 0, \frac{1}{3} - \sqrt{\frac{1}{3} \cdot 0}\Big \rbrace 
= \frac{2}{3} \approx 0.667.
\end{equation}

\subsection{Asymmetric $|\overline{W_3^L}\rangle$ State}

The asymmetric isosceles state is defined as $|\overline{W_3^L}\rangle = \frac{1}{2}|110\rangle + \frac{1}{2}|101\rangle + \frac{1}{\sqrt{2}}|011\rangle$ \cite{Lohmayer2006}.

\textbf{Vertex-Base Pair ($\rho_{AB}$):} Tracing out peripheral qubit C yields the super-link:
\begin{equation}
\rho_{AB} = \operatorname{Tr}_C(|\overline{W_3^L}\rangle\langle\overline{W_3^L}|) = \begin{pmatrix}
	0 & 0 & 0 & 0 \\
	0 & \frac{1}{2} & \frac{\sqrt{2}}{4} & 0 \\
	0 & \frac{\sqrt{2}}{4} & \frac{1}{4} & 0 \\
	0 & 0 & 0 & \frac{1}{4}
\end{pmatrix}.
\end{equation}
Applying Eq.~\ref{eq:master_concurrence}:
\begin{equation}
\mathcal{C}_{VB}(0) = 2\max\! \Big\lbrace 0, \frac{\sqrt{2}}{4} - \sqrt{0 \cdot \frac{1}{4}}\Big\rbrace
= \frac{1}{\sqrt{2}} \approx 0.707.
\end{equation}

\textbf{Base-Base Pair ($\rho_{BC}$):} Tracing out central vertex qubit A yields the peripheral link:
\begin{equation}
\rho_{BC} = \operatorname{Tr}_A(|\overline{W_3^L}\rangle\langle\overline{W_3^L}|) = \begin{pmatrix}
0 & 0 & 0 & 0 \\
0 & \frac{1}{4} & \frac{1}{4} & 0 \\
0 & \frac{1}{4} & \frac{1}{4} & 0 \\
0 & 0 & 0 & \frac{1}{2}
\end{pmatrix}.
\end{equation}
Applying Eq.~\ref{eq:master_concurrence}:
\begin{equation}
\mathcal{C}_{BB}(0) = 2\max\!\Big \lbrace 0, \frac{1}{4} - \sqrt{0 \cdot \frac{1}{2}}\Big\rbrace
= \frac{1}{2} = 0.500.
\end{equation}

This establishes the undecohered baseline hierarchy: $\mathcal{C}_{VB} > \mathcal{C}_W > \mathcal{C}_{BB}$.

\section{Analytical Derivations of Decoherence Dynamics}
\label{app:C}

\subsection{Phase Damping Dynamics}

Phase damping models the loss of quantum coherence without energy exchange \cite{Nielsen2010}. The single-qubit Kraus operators are given in Appendix~\ref{app:A}. We construct the two-qubit operators via $E_k = I \otimes K_k$ and apply them to the target subsystem.

\subsubsection{Symmetric $|W\rangle$ State}

Applying the phase damping channel to $\rho_{BC}^{(W)}$ yields a new X-state where the populations remain unchanged, but the off-diagonal coherence is suppressed by $\sqrt{1-p}$:
\begin{equation}
\rho'_{23} = \frac{\sqrt{1-p}}{3}, \quad \rho'_{11} = 0, \quad \rho'_{44} = 0.
\end{equation}
Applying the X-state formula (Eq.~\ref{eq:master_concurrence}):
\begin{equation}
\mathcal{C}_W(p) = 2\max\!\Big\lbrace 0, \frac{\sqrt{1-p}}{3}\Big\rbrace  = \frac{2}{3}\sqrt{1-p}.
\end{equation}

\subsubsection{Asymmetric State: Vertex-Base Pair}

Applying the channel locally to the base qubit $B$ in the $\rho_{AB}$ subsystem yields:
\begin{equation}
\rho'_{23} = \tfrac{\sqrt{2}}{4}\sqrt{1-p}, \quad \rho'_{11} = 0, \quad \rho'_{44} = \tfrac{1}{4}.
\end{equation}
Applying the X-state formula:
\begin{equation}
\mathcal{C}_{VB}(p) = 2\max\!\Big\lbrace 0, \frac{\sqrt{2(1-p)}}{4}\Big\rbrace
= \frac{1}{\sqrt{2}}\sqrt{1-p}.
\end{equation}

\subsubsection{Asymmetric State: Base-Base Pair}

Applying the channel locally to the base qubit $C$ in the $\rho_{BC}$ subsystem yields:
\begin{equation}
\rho'_{23} = \frac{1}{4}\sqrt{1-p}, \quad \rho'_{11} = 0, \quad \rho'_{44} = \frac{1-p}{2}.
\end{equation}
Applying the X-state formula:
\begin{equation}
\mathcal{C}_{BB}(p) = 2\max\!\Big\lbrace 0, \frac{\sqrt{1-p}}{4}\Big\rbrace = \frac{1}{2}\sqrt{1-p}.
\end{equation}

\subsection{Amplitude Damping Dynamics}

Amplitude damping models energy dissipation to a zero-temperature environment with decay probability $\gamma \in [0,1]$ \cite{Plenio1998}. The Kraus operators are given in Appendix~\ref{app:A}.

\subsubsection{Symmetric $|W\rangle$ State}

Applying the channel to $\rho_{BC}^{(W)}$ yields:
\begin{equation}
\rho'_{23} = \frac{\sqrt{1-\gamma}}{3}, \quad \rho'_{11} = \frac{\gamma}{3}, \quad \rho'_{44} = 0.
\end{equation}
Applying the X-state formula:
\begin{equation}
\mathcal{C}_W(\gamma) = 2\max\!\Big\lbrace 0, \frac{\sqrt{1-\gamma}}{3}\Big\rbrace = \frac{2}{3}\sqrt{1-\gamma}.
\end{equation}

\subsubsection{Asymmetric State: Vertex-Base Pair}

Applying the channel to $\rho_{AB}$ yields:
\begin{equation}
\rho'_{23} = \frac{\sqrt{2(1-\gamma)}}{4}, \quad
\rho'_{11} = \frac{\gamma}{2}, \quad \rho'_{44} = \frac{1-\gamma}{4}.
\end{equation}
Applying the X-state formula:
\begin{eqnarray}
\mathcal{C}_{VB}(\gamma) &=& 2\max\!\Big\lbrace 0,\, \frac{\sqrt{2(1-\gamma)}}{4} - \sqrt{\frac{\gamma(1-\gamma)}{8}}\Big\rbrace \nonumber\\
&=& \frac{\sqrt{1-\gamma}}{\sqrt{2}}\left(1 - \sqrt{\gamma}\right).
\end{eqnarray}

\subsubsection{Asymmetric State: Base-Base Pair}

Applying the channel to $\rho_{BC}$ yields:
\begin{equation}
\rho'_{23} = \frac{\sqrt{1-\gamma}}{4}, \quad
\rho'_{11} = \frac{\gamma}{4}, \quad \rho'_{44} = \frac{1-\gamma}{2}.
\end{equation}
Applying the X-state formula reveals the entanglement sudden death (ESD) term:
\begin{eqnarray}
\mathcal{C}_{BB}(\gamma) &=& 2\max\!\left(0,\, \frac{\sqrt{1-\gamma}}{4} - \sqrt{\frac{\gamma(1-\gamma)}{8}}\right)\nonumber\\
&=& \frac{\sqrt{1-\gamma}}{2}\left(1 - \sqrt{2\gamma}\right).
\end{eqnarray}

\subsection{Depolarization Dynamics}

Depolarization is modelled as isotropic noise with probability $p$ \cite{Nielsen2010}. For local noise on bipartite subsystems, the equivalent map is $\mathcal{E}_C(\rho) = (1{-}\lambda)\rho + \lambda \bigl(\operatorname{Tr}_C(\rho) \otimes \tfrac{I}{2}\bigr)$, where $\lambda = \frac{4}{3}p$.

\subsubsection{Symmetric $|W\rangle$ State}

Applying the depolarizing map to $\rho_{BC}^{(W)}$ yields:
\begin{equation}
\rho'_{23} = \frac{3-4p}{9}, \quad \rho'_{11} = \frac{1}{3}, \quad \rho'_{44} = \frac{2p}{9}.
\end{equation}
Applying the X-state formula:
\begin{eqnarray}
\mathcal{C}_W(p) &=& 2\max\!\Big\lbrace 0, \frac{|3-4p|}{9} - \frac{\sqrt{6p}}{9}\Big\rbrace \\
&=& \max\!\Big\lbrace 0, \frac{2}{9}\left(3-4p - \sqrt{6p}\right)\Big\rbrace.
\end{eqnarray}

\subsubsection{Asymmetric State: Vertex-Base Pair}

Applying the map to $\rho_{AB}$ yields:
\begin{equation}
\rho'_{23} = \frac{\sqrt{2}(3-4p)}{12}, \quad
\rho'_{11} = \frac{p}{3}, \quad \rho'_{44} = \frac{1}{4}.
\end{equation}
Applying the X-state formula:
\begin{eqnarray}
\mathcal{C}_{VB}(p) &=& 2\max\!\Big \lbrace 0, \frac{\sqrt{2}|3-4p|}{12} - \frac{\sqrt{3p}}{6}\Big \rbrace \nonumber\\
&=& \max\!\Big\lbrace 0, \frac{\sqrt{2}}{6}\left(|3-4p| - \sqrt{6p}\right)\Big\rbrace.
\end{eqnarray}

\subsubsection{Asymmetric State: Base-Base Pair}

Applying the map to $\rho_{BC}$ yields:
\begin{equation}
\rho'_{23} = \frac{3-4p}{12}, \quad
\rho'_{11} = \frac{p}{6}, \quad \rho'_{44} = \frac{3-p}{6}.
\end{equation}
Applying the X-state formula:
\begin{eqnarray}
\mathcal{C}_{BB}(p) &=& 2\max\! \Big\lbrace 0, \frac{|3-4p|}{12} - \frac{\sqrt{p(3-p)}}{6}\Big\rbrace \nonumber\\
&=&\max\Big\lbrace 0, \frac{1}{6}\left(|3-4p| - 2\sqrt{3p-p^2}\right)\Big\rbrace.
\end{eqnarray}

\subsection{Generalized Amplitude Damping Dynamics}

The Generalized Amplitude Damping Channel (GADC) models a thermal reservoir parameterized by interaction strength $p$ and temperature weights $\alpha$ (damping) and $\beta=1-\alpha$ (amplification) \cite{Srikanth2008}. The Kraus operators are given in Appendix~\ref{app:A}.

\subsubsection{Symmetric $|W\rangle$ State}

Summing the four Kraus contributions yields the decohered elements:
\begin{equation}
\rho'_{23} = \frac{\sqrt{1-p}}{3}, \quad
\rho'_{11} = \frac{1-p+2\alpha p}{3}, \quad \rho'_{44} = \frac{(1-\alpha)p}{3}.
\end{equation}
Applying the X-state formula:
\begin{align}
\mathcal{C}_W(p,\alpha) = \max\!\Big\lbrace 0,\,\tfrac{2}{3}\bigl[&\sqrt{1-p}\nonumber\\
& -\sqrt{(1-p+2\alpha p)(1-\alpha)p}\,\bigr]\Big\rbrace
\end{align}

\subsubsection{Asymmetric State: Vertex-Base Pair}

Applying the GADC locally yields:
\begin{equation}
\rho'_{23} = \frac{\sqrt{2(1-p)}}{4}, \quad
\rho'_{11} = \frac{\alpha p}{2}, \quad \rho'_{44} = \frac{1+p-2\alpha p}{4}.
\end{equation}
Applying the X-state formula:
\begin{align}
\mathcal{C}_{VB}(p,\alpha) =
\max\!\Big\lbrace 0,\,\tfrac{1}{\sqrt{2}}\bigl[\sqrt{1-p}
-\sqrt{\alpha p(1+p-2\alpha p)}\,\bigr]\Big\rbrace.
\end{align}

\subsubsection{Asymmetric State: Base-Base Pair}

Applying the GADC locally yields:
\begin{equation}
\rho'_{23} = \frac{\sqrt{1-p}}{4}, \quad
\rho'_{11} = \frac{\alpha p}{4}, \quad \rho'_{44} = \frac{2+p-3\alpha p}{4}.
\end{equation}
Applying the X-state formula, we obtain
\begin{align}
\mathcal{C}_{BB}(p,\alpha) = \max\!\Big\lbrace 0,\,\tfrac{1}{2}\bigl[\sqrt{1-p}
-\sqrt{\alpha p(2+p-3\alpha p)}\,\bigr]\Big\rbrace.
\end{align}
\newpage
\bibliographystyle{unsrt}
\addcontentsline{toc}{section}{References}
\bibliography{references.bib}

\end{document}